\documentclass[twocolumn,floatfix,superscriptaddress,showpacs]{revtex4}
\usepackage{graphicx}
\usepackage{amsmath}
\usepackage{bm}
\newcommand{\bra}[1]{\left\langle#1\right|}
\newcommand{\ket}[1]{\left|#1\right>}
\newcommand{\braket}[2]{\left\langle#1|#2\right\rangle}
\begin{document}

\title{Electronic shot noise in fractal conductors}
\author{C. W. Groth}
\affiliation{Instituut-Lorentz, Universiteit Leiden, P.O. Box 9506, 2300 RA
  Leiden, The Netherlands}
\author{J. Tworzyd{\l}o}
\affiliation{Institute of Theoretical Physics, University of Warsaw, Ho\.{z}a
  69, 00--681 Warsaw, Poland}
\author{C. W. J. Beenakker}
\affiliation{Instituut-Lorentz, Universiteit Leiden, P.O. Box 9506, 2300 RA
  Leiden, The Netherlands}
\date{February 2008}
\begin{abstract}
  By solving a master equation in the Sierpi\'{n}ski lattice and in a planar
  random-resistor network, we determine the scaling with size $L$ of the shot
  noise power $P$ due to elastic scattering in a fractal conductor. We find a
  power-law scaling $P\propto L^{d_{f}-2-\alpha}$, with an exponent depending
  on the fractal dimension $d_{f}$ and the anomalous diffusion exponent
  $\alpha$. This is the same scaling as the time-averaged current $\bar{I}$,
  which implies that the Fano factor $F=P/2e\bar{I}$ is scale independent. We
  obtain a value $F=1/3$ for anomalous diffusion that is the same as for normal
  diffusion, even if there is no smallest length scale below which the normal
  diffusion equation holds. The fact that $F$ remains fixed at $1/3$ as one
  crosses the percolation threshold in a random-resistor network may explain
  recent measurements of a doping-independent Fano factor in a graphene flake.
\end{abstract}
\pacs{73.50.Td, 05.40.Ca, 64.60.ah,  64.60.al}
\maketitle

Diffusion in a medium with a fractal dimension is characterized by an anomalous
scaling with time $t$ of the root-mean-squared displacement $\Delta$. The usual
scaling for integer dimensionality $d$ is $\Delta\propto t^{1/2}$, independent
of $d$. If the dimensionality $d_{f}$ is noninteger, however, an anomalous
scaling
\begin{equation}
\Delta\propto t^{1/(2+\alpha)}\label{alpha}
\end{equation}
with $\alpha>0$ may appear. This anomaly was discovered in the early 1980's
\cite{Web81,Ale82,Ben82,Gef83,Ram83} and has since been studied extensively
(see Refs.\ \cite{Hav87,Isi92} for reviews). Intuitively, the slowing down of
the diffusion can be understood as arising from the presence of obstacles at
all length scales---characteristic of a selfsimilar fractal geometry.

A celebrated application of the theory of fractal diffusion is to the scaling
of electrical conduction in random-resistor networks (reviewed in Refs.\
\cite{Sta94,Red07}). According to Ohm's law, the conductance $G$ should scale
with the linear size $L$ of a $d$-dimensional network as $G\propto L^{d-2}$. In
a fractal dimension the scaling is modified to $G\propto L^{d_{f}-2-\alpha}$,
depending both on the fractal dimensionality $d_{f}$ and on the anomalous
diffusion exponent $\alpha$. At the percolation threshold, the known
\cite{Hav87} values for $d=2$ are $d_{f}=91/48$ and $\alpha=0.87$, leading to a
scaling $G\propto L^{-0.97}$. This almost inverse-linear scaling of the
conductance of a planar random-resistor network contrasts with the
$L$-independent conductance $G\propto L^{0}$ predicted by Ohm's law in two
dimensions.

All of this body of knowledge applies to classical resistors, with applications
to disordered semiconductors and granular metals \cite{Shk84,Bun96}. The
quantum Hall effect provides one quantum mechanical realization of a
random-resistor network \cite{Tru83}, in a rather special way because
time-reversal symmetry is broken by the magnetic field. Very recently
\cite{Che07}, Cheianov, Fal'ko, Altshuler, and Aleiner announced an altogether
different quantum realization in zero magnetic field. Following experimental
\cite{Mar07} and theoretical \cite{Hwa07} evidence for electron and hole
puddles in undoped graphene \cite{note1}, Cheianov et al.\ modeled this system
by a degenerate electron gas in a random-resistor network. They analyzed both
the high-temperature classical resistance, as well as the low-temperature
quantum corrections, using the anomalous scaling laws in a fractal geometry.

These very recent experimental and theoretical developments open up new
possibilities to study quantum mechanical aspects of fractal diffusion, both
with respect to the Pauli exclusion principle and with respect to quantum
interference (which are operative in distinct temperature regimes). To access
the effect of the Pauli principle one needs to go beyond the time-averaged
current $\bar{I}$ (studied by Cheianov et al.\ \cite{Che07}), and consider the
time-dependent fluctuations $\delta I(t)$ of the current in response to a
time-independent applied voltage $V$. These fluctuations exist because of the
granularity of the electron charge, hence their name ``shot noise'' (for
reviews, see Refs.\ \cite{Bla00,Bee03}). Shot noise is quantified by the noise
power
\begin{equation}
P=2\int_{-\infty}^{\infty}dt\,\langle\delta I(0)\delta I(t)\rangle\label{Pdef}
\end{equation}
and by the Fano factor $F=P/2e\bar{I}$. The Pauli principle enforces $F<1$,
meaning that the noise power is smaller than the Poisson value
$2e\bar{I}$---which is the expected value for independent particles (Poisson
statistics).

The investigation of shot noise in a fractal conductor is particularly timely
in view of two different experimental results \cite{DiC07,Dan07} that have been
reported recently. Both experiments measure the shot noise power in a graphene
flake and find $F<1$. A calculation \cite{Two06} of the effect of the Pauli
principle on the shot noise of undoped graphene predicted $F=1/3$ in the
absence of disorder, with a rapid suppression upon either \textit{p}-type or
\textit{n}-type doping. This prediction is consistent with the experiment of
Danneau et al.\ \cite{Dan07}, but the experiment of DiCarlo et al.\
\cite{DiC07} gives instead an approximately \textit{doping-independent} $F$
near $1/3$. Computer simulations \cite{San07,Lew07} suggest that disorder in
the samples of DiCarlo et al.\ might cause the difference.

Motivated by this specific example, we study here the fundamental problem of
shot noise due to anomalous diffusion in a fractal conductor. While
\textit{equilibrium} thermal noise in a fractal has been studied previously
\cite{Ram84,Han86,Fou86}, it remains unknown how anomalous diffusion might
affect the \textit{nonequilibrium} shot noise. Existing studies
\cite{Kuz00,Cam03,Kin06} of shot noise in a percolating network were in the
regime where \textit{inelastic} scattering dominates, leading to hopping
conduction, while for diffusive conduction we need predominantly
\textit{elastic} scattering.

We demonstrate that anomalous diffusion affects $P$ and $\bar{I}$ in such a way
that the Fano factor (their ratio) becomes scale independent as well as
independent of $d_{f}$ and $\alpha$. Anomalous diffusion, therefore, produces
the same Fano factor $F=1/3$ as is known \cite{Bee92,Nag92} for normal
diffusion. This is a remarkable property of diffusive conduction, given that
hopping conduction in a percolating network does not produce a
scale-independent Fano factor \cite{Kuz00,Cam03,Kin06}. Our general findings
are consistent with the doping independence of the Fano factor in disordered
graphene observed by DiCarlo et al.\ \cite{DiC07}.

To arrive at these conclusions we work in the experimentally relevant regime
where the temperature $T$ is sufficiently high that the phase coherence length
is $\ll L$, and sufficiently low that the inelastic length is $\gg L$. Quantum
interference effects can then be neglected, as well as inelastic scattering
events. The Pauli principle remains operative if the thermal energy $kT$
remains well below the Fermi energy, so that the electron gas remains
degenerate.

We first briefly consider the case that the anomalous diffusion on long length
scales is preceded by normal diffusion on short length scales. This would
apply, for example, to a percolating cluster of electron and hole puddles with
a mean free path $l$ which is short compared to the typical size $a$ of a
puddle. We can then rely on the fact that $F=1/3$ for a conductor of any shape,
provided that the normal diffusion equation holds locally \cite{Naz94,Suk98},
to conclude that the transition to anomalous diffusion on long length scales
must preserve the one-third Fano factor.

This simple argument cannot be applied to the more typical class of fractal
conductors in which the normal diffusion equation does not hold on short length
scales. As representative for this class, we consider fractal lattices of sites
connected by tunnel barriers. The local tunneling dynamics then crosses over
into global anomalous diffusion, without an intermediate regime of normal
diffusion.

\begin{figure}[tb]
\centerline{\includegraphics[width=0.8\linewidth]{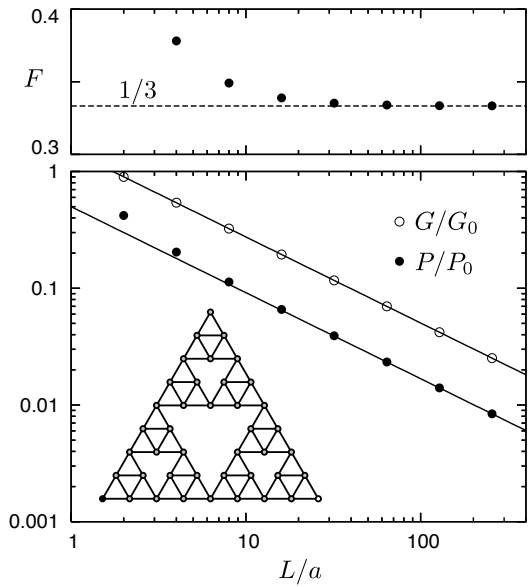}}
\caption{\label{fig_Sier} Lower panel: Electrical conduction through a
  Sierpi\'{n}ski lattice. This is a deterministic fractal, constructed by
  recursively removing a central triangular region from an equilateral
  triangle. The recursion level $r$ quantifies the size $L=2^{r}a$ of the
  fractal in units of the elementary bond length $a$ (the inset shows the
  fourth recursion). The conductance $G=\bar{I}/V$ (open dots, normalized by
  the tunneling conductance $G_{0}$ of a single bond) and shot noise power $P$
  (filled dots, normalized by $P_{0}=2eVG_{0}$) are calculated for a voltage
  difference $V$ between the lower-left and lower-right corners of the lattice.
  Both quantities scale as $L^{d_{f}-2-\alpha}=L^{\log_{2}(3/5)}$ (solid lines
  on the double-logarithmic plot). The Fano factor
  $F=P/2e\bar{I}=(P/P_{0})(G_{0}/G)$ rapidly approaches $1/3$, as shown in the
  upper panel.  }
\end{figure}

A classic example is the Sierpi\'{n}ski lattice \cite{Sie15} shown in Fig.\
\ref{fig_Sier} (inset). Each site is connected to four neighbors by bonds that
represent the tunnel barriers, with equal tunnel rate $\Gamma$ through each
barrier.  The fractal dimension is $d_{f}=\log_{2} 3$ and the anomalous
diffusion exponent is \cite{Hav87} $\alpha=\log_{2} (5/4)$. The Pauli exclusion
principle can be incorporated as in Ref.\ \cite{Liu97}, by demanding that each
site is either empty or occupied by a single electron. Tunneling is therefore
only allowed between an occupied site and an adjacent empty site. A current is
passed through the lattice by connecting the lower left corner to a source
(injecting electrons so that the site remains occupied) and the lower right
corner to a drain (extracting electrons so that the site remains empty). The
resulting stochastic sequence of current pulses is the ``tunnel exclusion
process'' of Ref.\ \cite{Roc05}.

The statistics of the current pulses can be obtained exactly (albeit not in
closed form) by solving a master equation \cite{Bag03}. We have calculated the
first two cumulants by extending to a two-dimensional lattice the
one-dimensional calculation of Ref.\ \cite{Roc05}. To manage the added
complexity of an extra dimension we found it convenient to use the Hamiltonian
formulation of Ref.\ \cite{Sch01}. The hierarchy of linear equations that we
need to solve in order to obtain $\bar{I}$ and $P$ is derived in the
appendix.

\begin{figure}[tb]
\centerline{\includegraphics[width=0.8\linewidth]{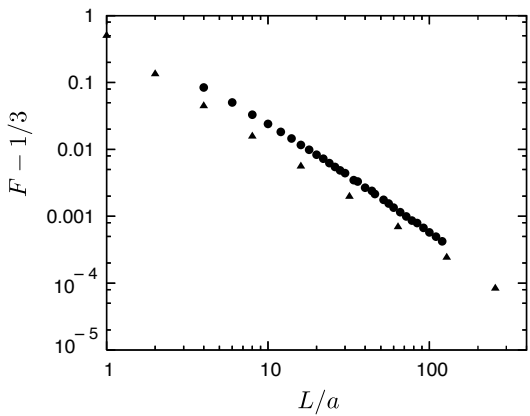}}
\caption{\label{fig_Fano} The deviation of the Fano factor from $1/3$ scales to
  zero as a power law for the Sierpi\'{n}ski lattice (triangles) and for the
  random-resistor network (circles).  }
\end{figure}

The results in Fig.\ \ref{fig_Sier} demonstrate, firstly, that the shot noise
power $P$ scales as a function of the size $L$ of the lattice with the same
exponent $d_{f}-2-\alpha=\log_{2}(3/5)$ as the conductance; and, secondly, that
the Fano factor $F$ approaches $1/3$ for large $L$. More precisely, see Fig.\
\ref{fig_Fano}, we find that $F-1/3\propto L^{-1.5}$ scales to zero as a power
law, with $F-1/3<10^{-4}$ for our largest $L$.

\begin{figure}[tb]
\centerline{\includegraphics[width=0.8\linewidth]{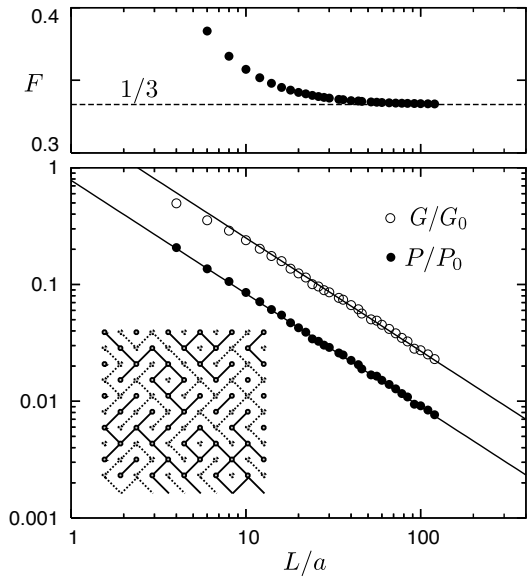}}
\caption{\label{fig_puddles} Same as Fig.\ \ref{fig_Sier}, but now for the
  random-resistor network of disordered graphene introduced by Cheianov et al.\
  \cite{Che07}. The inset shows one realization of the network for $L/a=10$
  (the data points are averaged over $\simeq 10^{3}$ such realizations). The
  alternating solid and dashed lattice sites represent, respectively, the
  electron (\textit{n}) and hole (\textit{p}) puddles. Horizontal bonds (not
  drawn) are \textit{p-n} junctions, with a negligibly small conductance
  $G_\textit{pn}\approx 0$. Diagonal bonds (solid and dashed lines) each have
  the same tunnel conductance $G_{0}$. Current flows from the left edge of the
  square network to the right edge, while the upper and lower edges are
  connected by periodic boundary conditions. This plot is for undoped graphene,
  corresponding to an equal fraction of solid (\textit{n-n}) and dashed
  (\textit{p-p}) bonds.  }
\end{figure}

Turning now to the application to graphene mentioned in the introduction, we
have repeated the calculation of shot noise and Fano factor for the
random-resistor network of electron and hole puddles introduced by Cheianov et
al.\ \cite{Che07}. The results, shown in Fig.\ \ref{fig_puddles}, demonstrate
that the shot noise power $P$ scales with the same exponent $L^{-0.97}$ as the
conductance $G$ (solid lines in the lower panel), and that the Fano factor $F$
approaches $1/3$ for large networks (upper panel). This is a random, rather
than a deterministic fractal, so there remains some statistical scatter in the
data, but the deviation of $F$ from $1/3$ for the largest lattices is still
$<10^{-3}$ (see the circular data points in Fig.\ \ref{fig_Fano}).

In conclusion, we have found that the universality of the one-third Fano
factor, previously established for normal diffusion
\cite{Bee92,Nag92,Naz94,Suk98}, extends to anomalous diffusion as well. This
universality might have been expected with respect to the fractal dimension
$d_{f}$ (since the Fano factor is dimension independent), but we had not
expected universality with respect to the anomalous diffusion exponent
$\alpha$. The experimental implication of the universality is that the Fano
factor remains fixed at $1/3$ as one crosses the percolation threshold in a
random-resistor network---thereby crossing over from anomalous diffusion to
normal diffusion. This is consistent with the doping-independent Fano factor
measured in a graphene flake by DiCarlo et al.\ \cite{DiC07}.

A discussion with L. S. Levitov motivated this research, which was supported by
the Netherlands Science Foundation NWO/FOM.  We also acknowledge support by the
European Community's Marie Curie Research Training Network under
contract MRTN-CT-2003-504574, Fundamentals of Nanoelectronics.

\appendix
\section{Calculation of the Fano factor for the tunnel exclusion process on a
  two-dimensional network}

Here we present the method we used to calculate the Fano factor for the tunnel
exclusion process in the Sierpi\'{n}ski lattice and in the random-resistor
network. We follow the master equation approach of Refs.\ \cite{Roc05,Bag03}.
The two-dimensionality of our networks requires a more elaborate bookkeeping,
which we manage by means of the Hamiltonian formalism of Ref.\ \cite{Sch01}.

\subsection{Counting statistics}

We consider a network of $N$ sites, each of which is either empty or singly
occupied. Two sites are called adjacent if they are directly connected by at
least one bond.  A subset $\mathcal S$ of the $N$ sites is connected to the
source and a subset $\mathcal D$ is connected to the drain. Each of the $2^N$
possible states of the network is reached with a certain probability at time
$t$. We store these probabilities in the $2^N$-dimensional vector $\ket{P(t)}$.
Its time evolution in the tunnel exclusion process is given by the master
equation
\begin{equation}
  \label{eq:evol_Pt}
  \frac{d}{d t}\ket{P(t)} = M \ket{P(t)},
\end{equation}
where the matrix $M$ contains the tunnel rates.  The normalization condition
can be written as $\braket{\Sigma}{P}=1$, in terms of a vector $\bra{\Sigma}$
that has all $2^{N}$ components equal to 1. This vector is a left eigenstate of
$M$ with zero eigenvalue
\begin{equation}
\bra{\Sigma}M=0,\label{sM0}
\end{equation}
because every column of $M$ must sum to zero in order to conserve probability.
The right eigenstate with zero eigenvalue is the stationary distribution
$\ket{P_\infty}$. All other eigenvalues of $M$ have a real part $<0$.

We store in the vector $\ket{P(t,Q)}$ the conditional probabilities that a
state is reached at time $t$ after precisely $Q$ charges have entered the
network from the source. Because the source remains occupied, a charge which
has entered the network cannot return back to the source but must eventually
leave through the drain. One can therefore use $Q$ to represent the number of
transfered charges. The time evolution of $\ket{P(t,Q)}$ reads
\begin{equation}
  \label{eq:evol_PtQ}
  \frac{d}{d t}\ket{P(t,Q)} =  M_0 \ket{P(t,Q)}
  + M_1 \ket{P(t,Q-1)},
\end{equation}
where $M = M_0 + M_1$ has been decomposed into a matrix $M_0$
containing all transitions by which $Q$ does not change and a matrix $M_1$
containing all transitions that increase $Q$ by 1.

The probability $\braket{\Sigma}{P(t,Q)}$ that $Q$ charges have been
transferred through the network at time $t$ represents the counting statistics.
It describes the entire statistics of current fluctuations. The cumulants
\begin{equation}
  \label{eq:cumulants}
  C_n = \left. \frac{\partial^n S(t,\chi)}{\partial \chi^n} \right|_{\chi=0}
\end{equation}
are obtained from the cumulant generating function
\begin{equation}
  \label{eq:cgf}
  S(t, \chi) = \ln \left[ \sum_Q \braket{\Sigma}{P(t,Q)} e^{\chi Q} \right].
\end{equation}
The average current and Fano factor are given by
\begin{equation}
\bar{I} =\lim_{t\rightarrow\infty} C_1 / t,\;\; F
= \lim_{t\rightarrow\infty} C_2 / C_1.\label{IFdef}
\end{equation}

The cumulant generating function \eqref{eq:cgf} can be expressed in terms of a
Laplace transformed probability vector $\ket{P(t,\chi)} = \sum_Q \ket{P(t,Q}
e^{\chi Q}$ as
\begin{equation}
  \label{eq:cgf2}
  S(t, \chi) = \ln \braket{\Sigma}{P(t,\chi)}.
\end{equation}
Transformation of Eq.\ \eqref{eq:evol_PtQ} gives
\begin{equation}
  \label{eq:evol_Ptx}
  \frac{d}{d t} \ket{P(t,\chi)} = M(\chi) \ket{P(t,\chi)},
\end{equation}
where we have introduced the counting matrix
\begin{equation}
  \label{eq:general_counting_matrix}
  M(\chi) = M_0 + e^\chi M_1 .
\end{equation}
The cumulant generating function follows from
\begin{equation}
S(t, \chi) =
  \ln \bra{\Sigma} e^{t M(\chi)}\ket{P(0, \chi)}.\label{Schidef}
\end{equation}

The long-time limit of interest for the Fano factor can be implemented as
follows \cite{Bag03}. Let $\mu(\chi)$ be the eigenvalue of $M(\chi)$ with the
largest real part, and let $\ket{P_\infty(\chi)}$ be the corresponding
(normalized) right eigenstate,
\begin{align}
  \label{eq:eigenstate}
 & M(\chi) \ket{P_\infty(\chi)} = \mu(\chi) \ket{P_\infty(\chi)},\\
 & \braket{\Sigma}{P_{\infty}(\chi)}=1.\label{Pinftynorm}
\end{align}
Since the largest eigenvalue of $M(0)$ is zero, we have
\begin{equation}
M(0)\ket{P_{\infty}(0)}=0\Leftrightarrow\mu(0)=0.\label{M0Pinfty}
\end{equation}
(Note that $\ket{P_{\infty}(0)}$ is the stationary distribution
$\ket{P_{\infty}}$ introduced earlier.) In the limit $t\rightarrow\infty$ only
the largest eigenvalue contributes to the cumulant generating function,
\begin{equation}
  \label{eq:cgf_mu} \lim_{t\rightarrow\infty}\frac{1}{t}S(t, \chi) =
  \lim_{t\rightarrow\infty}\frac{1}{t} \ln\,
  [e^{t \mu(\chi)} \braket{\Sigma}{P_{\infty}(\chi)}] =\mu(\chi).
\end{equation}

\subsection{Construction of the counting matrix}

The construction of the counting matrix $M(\chi)$ is simplified by expressing
it in terms of raising and lowering operators, so that it resembles a
Hamiltonian of quantum mechanical spins \cite{Sch01}.  First, consider a single
site with the basis states $\ket{0} = {1\choose 0}$ (vacant) and $\ket{1} =
{0\choose 1}$ (occupied).  We define, respectively, raising and lowering
operators
\begin{equation}
  s^+ =
  \begin{pmatrix}
    0 & 0 \\ 1 & 0
  \end{pmatrix},
  \quad 
  s^- =
  \begin{pmatrix}
    0 & 1 \\ 0 & 0
  \end{pmatrix}.
\end{equation}
We also define the electron number operator $n = s^+ s^-$ and the hole number
operator $\nu = \openone - n$ (with $\openone$ the $2\times 2$ unit matrix).
Each site $i$ has such operators, denoted by $s^+_i$, $s^-_i$, $n_i$, and
$\nu_i$.  The matrix $M(\chi)$ can be written in terms of these operators as
\begin{align}
  \label{eq:our_counting_matrix}
  M(\chi) ={}&
  \sum_{\langle i,j\rangle} \left(s^+_j s^-_i - \nu_j n_i \right) \nonumber\\
  &+\sum_{i\in\mathcal{S}} (e^\chi s^+_i - \nu_i)
  +\sum_{i\in\mathcal{D}} (s^-_i - n_i),
\end{align}
where all tunnel rates have been set equal to unity.  The first sum runs over
all ordered pairs $\langle i,j\rangle$ of adjacent sites.  These are Hermitian
contributions to the counting matrix. The second sum runs over sites in
$\mathcal{S}$ connected to the source, and the third sum runs over sites in
$\mathcal{D}$ connected to the drain. These are non-Hermitian contributions.

It is easy to convince oneself that $M(0)$ is indeed $M$ of Eq.\
\eqref{eq:evol_Pt}, since every possible tunneling event corresponds to two
terms in Eq.\ \eqref{eq:our_counting_matrix}: one positive non-diagonal term
responsible for probability gain for the new state and one negative diagonal
term responsible for probability loss for the old state.  In accordance with
Eq.\ \eqref{eq:general_counting_matrix}, the full $M(\chi)$ differs from $M$ by
a factor $e^\chi$ at the terms associated with charges entering the network.

\subsection{Extraction of the cumulants}

In view of Eq.\ \eqref{eq:cgf_mu}, the entire counting statistics in the
long-time limit is determined by the largest eigenvalue $\mu(\chi)$ of the
operator \eqref{eq:our_counting_matrix}. However, direct calculation of that
eigenvalue is feasible only for very small networks.  Our approach, following
Ref.\ \cite{Roc05}, is to derive the first two cumulants by solving a hierarchy
of linear equations.

We define
\begin{align}
  \label{eq:Ti}
  &T_i = \bra{\Sigma} n_i \ket{P_\infty(\chi)} =
  1-\bra{\Sigma} \nu_i \ket{P_\infty(\chi)},\\
  \label{eq:Uij}
  & U_{ij} = U_{ji} = \bra{\Sigma} n_i n_j \ket{P_\infty(\chi)}\quad\textrm{for}\
  i\neq j,\\
  & U_{ii}=2T_{i}-1.\label{Uii}
\end{align}
The value $T_i|_{\chi=0}$ is the average stationary occupancy of site $i$.
Similarly, $U_{ij}|_{\chi=0}$ for $i\neq j$ is the two-point correlator.

We will now express $\mu(\chi)$ in terms of $T_i$. We start from the definition
\eqref{eq:eigenstate}. If we act with $\bra{\Sigma}$ on the left-hand-side of
Eq.\ \eqref{eq:eigenstate} we obtain
\begin{align}
  \label{eq:lhs}
  \bra{\Sigma}  M(0) &+ (e^\chi - 1) \sum_{i\in\mathcal{S}}
    s^+_i  \ket{P_\infty(\chi)}  \nonumber\\
    &=  (e^\chi - 1)
    \sum_{i\in\mathcal{S}} \bra{\Sigma} s^+_i \ket{P_\infty(\chi)} \nonumber\\
    &=(e^\chi - 1) \sum_{i\in\mathcal{S}} \bra{\Sigma}
    \nu_i \ket{P_\infty(\chi)}  \nonumber\\
    &=(e^\chi - 1) \sum_{i\in\mathcal{S}} (1 - T_i).
\end{align}
In the second equality we have used Eq.\ \eqref{sM0} [which holds since
$M\equiv M(0)$]. Acting with $\bra{\Sigma}$ on the the right-hand-side of Eq.\
\eqref{eq:eigenstate} we obtain just $\mu(\chi)$, in view of Eq.\
\eqref{Pinftynorm}. Hence we arrive at
\begin{equation}
  \label{eq:mu}
  \mu(\chi) = (e^\chi - 1) \sum_{i\in\mathcal{S}} (1-T_i).
\end{equation}

From Eq.\ \eqref{eq:mu} we obtain the average current and Fano factor in terms
of $T_{i}$ and the first derivative $T'_{i}=dT_{i}/d\chi$ at $\chi=0$,
\begin{align}
  \label{eq:current}
  &\bar{I} = \lim_{t\to\infty} C_1 / t =
  \mu'(0) = \sum_{i\in\mathcal{S}} (1-T_i|_{\chi=0}),
\\
  \label{eq:fano}
  &F = \lim_{t\to\infty} \frac{C_2}{C_1} = \frac{\mu''(0)}{\mu'(0)} =
  1 - \frac{2 \sum_{i\in\mathcal{S}} T'_i|_{\chi=0}}
  {\sum_{i\in\mathcal{S}} (1-T_i|_{\chi=0})}.
\end{align}

\subsubsection{Average current}

To obtain $T_{i}$ we set up a system of linear equations starting from
\begin{equation}
  \mu(\chi) T_i =
  \bra{\Sigma}n_i M(\chi)\ket{P_{\infty}(\chi)}.\label{muchiT}
\end{equation}
Commuting $n_i$ to the right, using the commutation relations $[n_i, s_i^+] =
s_i^+$ and $[n_i, s_i^-] = - s_i^-$, we find
\begin{equation}
  \label{eq:muTi}
  \mu(\chi) T_i = {\sum_{j(i)}}T_{j} - k_{i}T_i
  + k_{i,\mathcal{S}} + (e^\chi - 1)\sum_{l\in\mathcal{S}}(T_i - U_{li}).
\end{equation}
The notation $\sum_{j(i)}$ means that the sum runs over all sites $j$ adjacent
to $i$.  The number $k_i$ is the total number of bonds connected to site $i$;
$k_{i,\mathcal{S}}$ of these bonds connect site $i$ to the source.

In order to compute $T_i|_{\chi=0}$ we set $\chi = 0$ in Eq.\ \eqref{eq:muTi},
use Eq.\ \eqref{M0Pinfty} to set the left-hand-side to zero, and solve the
resulting symmetric sparse linear system of equations, \begin{equation} -
  k_{i,\mathcal{S}} = \sum_{j(i)} T_j - k_i T_i. \end{equation} This is the
first level of the hierarchy. Substitution of the solution into Eq.\
\eqref{eq:current} gives the average current $\bar{I}$.

\subsubsection{Fano  factor}
To calculate the Fano factor via Eq.\ \eqref{eq:fano} we also need
$T'_i|_{\chi=0}$.  We take Eq.\ \eqref{eq:muTi}, substitute Eq.\ \eqref{eq:mu}
for $\mu(\chi)$, differentiate and set $\chi = 0$ to arrive at
\begin{equation}
  \sum_{l\in\mathcal{S}} (U_{li} - T_l T_i) - k_{i,\mathcal{S}}
  = \sum_{j(i)} T_j' - k_i T_i'.
\end{equation}
To find $U_{ij}|_{\chi=0}$ we note that
\begin{equation}
  \mu(\chi) U_{ij}
  = \bra{\Sigma} n_i n_j M(\chi)\ket{P_{\infty}(\chi)},\;\;i\neq j,
\end{equation}
and commute $n_i$ to the right.  Setting $\chi = 0$ provides the second level
of the hierarchy of linear equations,
\begin{align}
  \label{eq:Uijnew}   0 ={}& \sum_{l(j),l\neq i} U_{il}
  + \sum_{l(i), l\neq j} U_{jl}   - (k_i + k_j - 2 d_{ij}) U_{ij}\nonumber\\
  &+k_{j,\mathcal{S}}T_{i}+k_{i,\mathcal{S}}T_{j},\;\;i\neq j.
\end{align}
The number $d_{ij}$ is the number of bonds connecting sites $i$ and $j$ if they
are adjacent, while $d_{ij}=0$ if they are not adjacent.


\begin{thebibliography}{99}
\bibitem{note2} An electron gas is called ``degenerate'' if the average
  occupation number of a quantum state is either close to unity or close to
  zero. It is called ``nondegenerate'' if the average occupation number is much
  smaller than unity for all states.
\end{thebibliography}
\end{document}